\documentclass[sigconf]{acmart}
\usepackage{booktabs}
\usepackage{amsmath}
\usepackage{xcolor}
\usepackage{xspace}
\usepackage{tcolorbox}
\usepackage[table]{xcolor}
\usepackage{subfigure}
\usepackage{fix-cm}
\usepackage{enumitem}
\usepackage{setspace}
\usepackage{tikz}
\usetikzlibrary{tikzmark}
\usepackage{float}
\usepackage{dsfont}
\usepackage{pifont}
\usepackage{bbding}
\usepackage{graphicx}
\usepackage{subfigure}
\usepackage{subcaption}
\usepackage{bbm}
\usepackage{multirow}
\usepackage{bigdelim}
\usepackage{hyperref}  

\definecolor{bluecolor}{RGB}{0,114,178} 
\definecolor{redcolor}{RGB}{213,94,0}   

\usepackage{soul}
\definecolor{pastelblue}{RGB}{173, 216, 230}
\definecolor{pastelred}{RGB}{255, 182, 193}
\definecolor{lightgrey}{RGB}{235,235,235} 
\definecolor{mediumgrey}{RGB}{215,215,215} 

\newcommand{\method}{PairSem\xspace}
\newcommand{\methodfast}{PairSem$_{\text{fast}}$\xspace}

\AtBeginDocument{%
  }

\setcopyright{acmlicensed}
\copyrightyear{2026}
\acmYear{2026}
\setcopyright{cc}
\setcctype{by}
\acmConference[WWW '26]{Proceedings of the ACM Web Conference 2026}{April 13--17, 2026}{Dubai, United Arab Emirates}
\acmBooktitle{Proceedings of the ACM Web Conference 2026 (WWW '26), April 13--17, 2026, Dubai, United Arab Emirates}
\acmPrice{}
\acmDOI{10.1145/3774904.3792657}
\acmISBN{979-8-4007-2307-0/2026/04}

\begin{document}

\title{PairSem: LLM-Guided Pairwise Semantic Matching\\for Scientific Document Retrieval} 

\author{Wonbin Kweon}
\affiliation{%
  \institution{University of Illinois \\ Urbana-Champaign}
  \city{Urbana}
  \state{IL}
  \country{USA}
}
\email{wonbin@illinois.edu}

\author{Runchu Tian}
\affiliation{%
  \institution{University of Illinois \\ Urbana-Champaign}
  \city{Urbana}
  \state{IL}
  \country{USA}
}
\email{runchut2@illinois.edu}

\author{SeongKu Kang}
\affiliation{%
  \institution{Korea University}
  \city{Seoul}
  \country{Republic of Korea}
}
\email{seongkukang@korea.ac.kr}

\author{Pengcheng Jiang}
\affiliation{%
  \institution{University of Illinois \\ Urbana-Champaign}
  \city{Urbana}
  \state{IL}
  \country{USA}
}
\email{pj20@illinois.edu}

\author{Zhiyong Lu}
\affiliation{%
  \institution{\mbox{National Institutes of Health}}
  \city{Bethesda}
  \state{MD}
  \country{USA}
}
\email{zhiyong.lu@nih.gov}

\author{Jiawei Han}
\affiliation{%
  \institution{University of Illinois \\ Urbana-Champaign}
  \city{Urbana}
  \state{IL}
  \country{USA}
}
\email{hanj@illinois.edu}

\author{Hwanjo Yu}
\affiliation{
    \institution{Pohang University of Science and Technology}
    \city{Pohang}
    \country{Republic of Korea}
}
\email{hwanjoyu@postech.ac.kr}

\begin{abstract}
Scientific document retrieval is a critical task for enabling knowledge discovery and supporting research across diverse domains.
However, existing dense retrieval methods often struggle to capture fine-grained scientific concepts in texts due to their reliance on holistic embeddings and limited domain understanding.
Recent approaches leverage large language models (LLMs) to extract fine-grained semantic entities and enhance semantic matching, but they typically treat entities as independent fragments, overlooking the multi-faceted nature of scientific concepts.
To address this limitation, we propose Pairwise Semantic Matching (\textbf{\method}), a framework that represents relevant semantics as entity–aspect pairs, capturing complex, multi-faceted scientific concepts.
\method is unsupervised, base retriever-agnostic, and plug-and-play, enabling precise and context-aware matching without requiring query-document labels or entity annotations.
Extensive experiments on multiple datasets and retrievers demonstrate that \method significantly improves retrieval performance, highlighting the importance of modeling multi-aspect semantics in scientific information retrieval.
\end{abstract}



\maketitle

\section{Introduction}
\begin{figure}[ht!]
  \centering
  \includegraphics[width=1\linewidth]{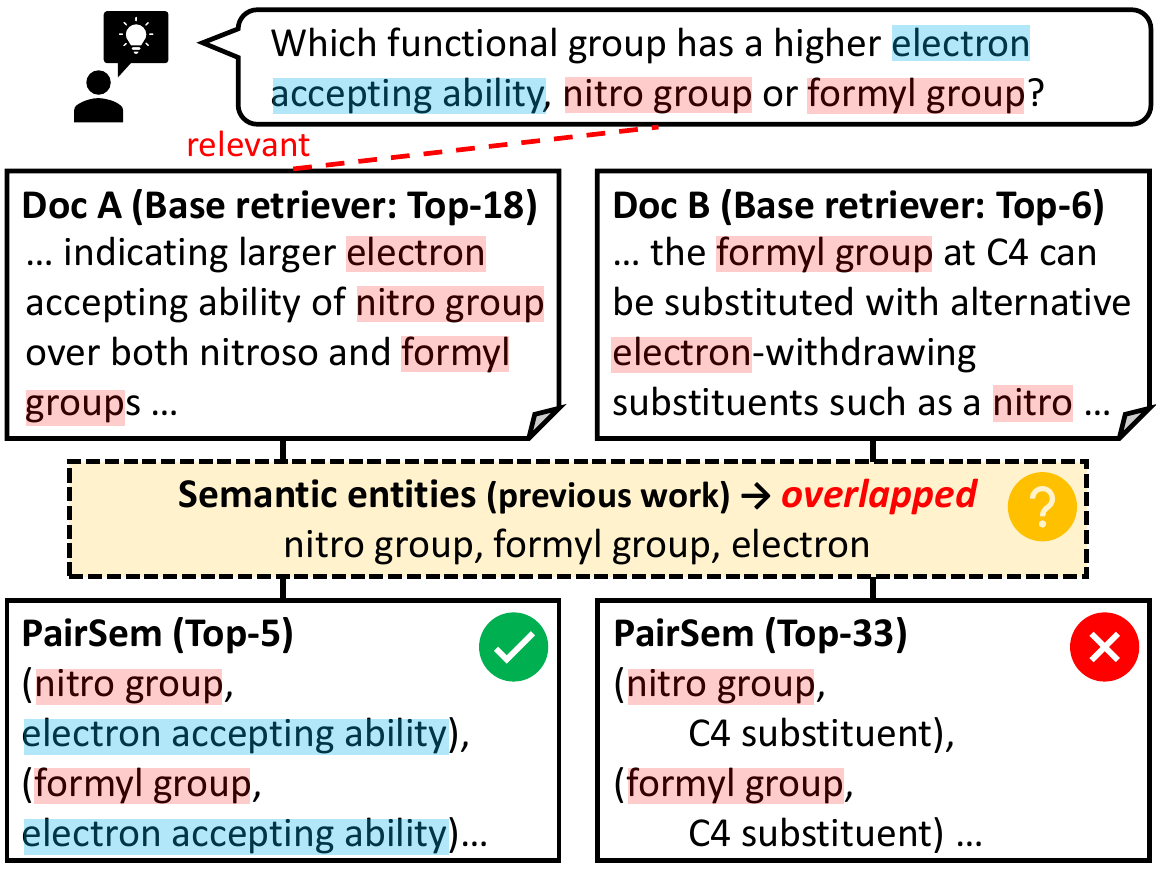}
  \vspace{-0.5cm}
  \caption{Example of document retrieval in Chemistry domain. We highlighted relevant entities in \sethlcolor{pastelred}\hl{Red} and associated aspects in \sethlcolor{pastelblue}\hl{Blue}.}  
  \vspace{-0.3cm}
\label{fig:intro}
\end{figure}

Scientific document retrieval plays a fundamental role in discovering academic knowledge and accelerating scientific research \cite{toter, semrank, cai2024mixgr, kang2025improving}.
Before exploring new ideas, researchers first need to identify relevant prior work to build upon existing findings.
To support this process, specialized web search services such as Google Scholar\footnote{https://scholar.google.com/} and PubMed\footnote{https://pubmed.ncbi.nlm.nih.gov/} provide access to vast collections of scientific literature.
Furthermore, with the growing use of large language models (LLMs) for scientific reasoning tasks \cite{chemrag, tang2025chemagent, ouyang2024structured}, retrieving precise and contextually relevant passages becomes crucial for supplying the rich background knowledge needed to generate accurate and insightful responses.

Dense retrieval \cite{dpr, xiong2020approximate, wang2022text} has recently emerged as the predominant approach for document search.
These models, often built on pre-trained language models, are fine-tuned to map queries and documents into a shared vector space, where relevance is measured by embedding similarity.
While effective in general web search \cite{dpr, specterv2, contriever}, their performance often falls short in scientific domains \cite{cai2024mixgr, kang2025improving}.
One limitation stems from pre-training on general-domain text, which provides limited understanding of specialized terminology and domain-specific user intent \cite{jin2023medcpt,xu2024bmretriever}.
Another challenge is that dense embeddings provide only holistic document-level representations, making it difficult to capture fine-grained details such as the target property of a specific chemical compound \cite{lmitationofemb, shavarani2025entity, chen2022salient}.
In addition, constructing the annotated query–document pairs required for fine-tuning is costly and relies heavily on domain expertise, limiting adaptability to specialized fields \cite{toter, semrank, shi2025hypercube, lee2026improving}.

Recent advances in LLMs \cite{gpt4o, claude3, llama3} have encouraged new approaches that generate rich semantics relevant to queries and documents, by using the powerful comprehension ability of LLMs.
Early methods prompted LLMs to generate pseudo queries \cite{HyDE}, keywords \cite{grf, boudin2020keyphrase}, or related sentences \cite{zheng2020bertqe,CSQE} to expand the original texts.
However, these approaches still encode the generated semantics together with the original texts into a single embedding, which limits the ability to capture fine-grained entities such as \texttt{"nitro group"} or \texttt{"formyl group"} in Figure~\ref{fig:intro}.
To address this, recent work \cite{toter, semrank, topick} builds a separate list of semantic entities (e.g., yellow box in Figure~\ref{fig:intro}), explicitly linking documents with scientific concepts drawn from the pre-defined set \cite{mag}. 
Rather than integrating the semantic entities into document embeddings, these methods incorporate an additional semantic matching score alongside the embedding-based score, enhancing retrieval in scientific domains.

While this semantic matching improves alignment between queries and documents, most existing methods treat scientific concepts as independent fragments.
This approach overlooks the multi-faceted characteristics of a scientific entity, where meaning often emerges from the interplay between entities and their associated aspects.
For instance, in Figure~\ref{fig:intro}, both documents mention \texttt{"nitro group"}, sharing most of semantic entities (yellow box).
However, Document A describes \texttt{"electron accepting ability"} (aspect) of \texttt{"nitro group"} (entity), while Document B highlights \texttt{"C4 substituent"} (another aspect) of the same entity.
By collapsing these diverse entity–aspect relationships into a single-level flat list, current methods struggle to capture fine-grained matches between queries and documents.
Indeed, our analysis demonstrates that entities are frequently associated with multiple aspects (6.75 on average in ChemLit-QA dataset).
This motivates the need for a new retrieval framework that explicitly models entities together with their multiple aspects, enabling more precise and context-aware matching in scientific document retrieval.

We propose LLM-Guided \textbf{Pair}wise \textbf{Sem}antic Matching (\textbf{\method}), a framework for fine-grained semantic matching between scientific queries and documents.  
The key idea is to represent semantics as entity-aspect pairs, capturing the multi-faceted nature of scientific concepts.  
Here, \textit{entities} represent scientific objects, such as chemical compounds (e.g., \texttt{"perfluorinated acid"}) or computer science models (e.g., \texttt{"gpt-oss-120b"}), while \textit{aspects} capture their corresponding properties or attributes, such as \texttt{"melting point"} or \texttt{"post-training"}, respectively.
In this work, we generate these pairs without supervision, eliminating the need for external entity sets or human annotations.
To ensure consistent terminology across documents, entity and aspect names are normalized through corpus-level set construction with synonym merging.
Then, pair generation is further augmented with candidate entities and aspects drawn from the constructed set, preventing LLMs from missing relevant pairs.
By representing queries and documents with these structured semantic pairs, \method enables precise and fine-grained matching, improving retrieval accuracy in scientific domains.

The proposed \method is a plug-and-play framework that can be seamlessly integrated with existing dense retrievers.
Importantly, \method enhances retrieval performance in scientific domains without relying on query-document relevance labels or document-entity annotations.
Moreover, to enable efficient inference, we introduce \textbf{PairSem$_{\text{fast}}$}, which replaces LLM-based query semantic generation with lightweight entity and aspect predictors.
The main contributions of this paper are as follows:
\begin{itemize}[leftmargin=10pt, itemsep=2pt]
    \item We identify and model the multi-faceted aspects of scientific entities, capturing complex semantic relationships that were previously overlooked in previous literature.
    \item We propose \method, a retriever-agnostic framework that generates entity-aspect pairs from texts, to enhance scientific document retrieval with pairwise semantic matching.
    \item We demonstrate the superiority of \method through extensive experiments using three base retrievers of different sizes and three datasets spanning diverse scientific domains.
\end{itemize}

\section{Problem Formulation}
Given a query $q$, a retriever returns a ranked list of documents $d \in \mathcal{D}$. 
Specifically, a dense retriever $\mathcal{R}(\cdot)$ encodes the query and documents separately into a shared embedding space, and relevance is measured by embedding similarity: 
\begin{equation}
    sim_{\text{base}}(q,d) = \textbf{e}_q^\top \textbf{e}_d,
\end{equation}
where $\textbf{e}_q = \mathcal{R}(q)$ and $\textbf{e}_d = \mathcal{R}(d)$ denote the embeddings of the query and the document, respectively. 
Document embeddings are pre-computed before the query is issued and efficiently retrieved using approximate nearest-neighbor (ANN) search methods \cite{faiss}. 
In this work, we focus on scientific document retrieval, where $\mathcal{D}$ consists of scientific documents such as research papers. 
We assume that the base retriever is pre-trained on a general corpus and \textit{not} fine-tuned on the target corpus (i.e., zero-shot retrieval). 
Our goal is to enhance the retrieval performance of the base retriever without relying on query-document relevance labels for the target corpus.

\section{Related Work}
\noindent \textbf{Pre-trained dense retrievers.}
Pre-trained dense retrievers map texts into continuous Euclidean spaces, where semantic relevance between queries and documents is captured through vector similarity. 
This paradigm has become fundamental in modern information retrieval (IR)~\citep{faloutsos1998survey, mei2000IR, mitra2000information} because of its state-of-the-art performance~\citep{zhao2024densesurvey, Hambarde2023IRsurvey}.
Early studies~\citep{dpr, ren-etal-2021-pair, xiong2020approximate, luan-etal-2021-sparse, zhan2021optimizingdenseretrievalmodel, qu-etal-2021-rocketqa, contriever} built upon general-domain pre-trained transformer models~\citep{attention} such as BERT~\citep{devlin-etal-2019-bert} and ERNIE~\citep{zhang-etal-2019-ernie}, and demonstrated the effectiveness of dense retrieval by fine-tuning these models on query-document pairs using various negative sampling strategies.
Subsequent works, such as SciBERT~\citep{beltagy-etal-2019-scibert}, SPECTER~\citep{specterv2}, and BMRetriever~\citep{xu2024bmretriever}, focused on scientific retrieval by fine-tuning retriever models on domain-specific data.
Recently, retrievers built upon large language models (LLMs)~\citep{gpt4o, yang2025qwen3, llama3}, such as GritLM~\citep{muennighoff2025generativerepresentationalinstructiontuning}, NV-embed~\citep{lee2025nvembedimprovedtechniquestraining}, EmbeddingGemma \cite{vera2025embeddinggemma}, and Qwen3-Embedding~\citep{yang2025qwen3}, have further leveraged their strong text understanding capabilities, substantially improving the performance of dense retrieval.

While effective in general-domain web search, the performance of pre-trained dense retrievers is often limited in scientific domains~\citep{toter, tian2025corankllmbasedcompactreranking}.
Two key challenges are (i) \emph{out-of-distribution (OOD)} and (ii) \emph{granularity}.
On the one hand, retrievers pre-trained on general-domain corpora lack specialized scientific knowledge, while scientific domain annotation is expensive and scarce~\citep{semrank}.
On the other hand, embeddings provide holistic representations at the document level, whereas scientific queries often require matching fine-grained details inside the documents, like a target property of a specific chemical compound~\citep{chen2022salient, shavarani2025entity}. 
PairSem mitigates these two limitations by extracting and matching entity-aspect pairs in documents and queries, enabling fine-grained matching in a training-free manner.

\vspace{+0.1cm}
\noindent \textbf{LLM-enhanced retrieval.} 
The emergence of LLMs~\citep{few-shot, chowdhery2022palmscalinglanguagemodeling, touvron2023llamaopenefficientfoundation} brought substantial advances to information retrieval.
The pretrained knowledge and instruction-following capabilities of LLMs can improve retrieval effectiveness without additional query–document relevance annotations \citep{kweon2025uncertainty}.

One line of work leverages LLMs as \emph{query rewriters} or \emph{expanders}, using their pretrained knowledge to refine queries or supplement background information.
Early works like ~\citep{boudin2020keyphrase} use language models to generate keyphrases for improving the generalization of document retrieval in scientific domain.
Rewrite-Retrieve-Read~\citep{ma-etal-2023-query} trains LLMs as query rewriters with reinforcement learning, which consistently improve retrieval accuracy in QA tasks.
GRF~\citep{grf} replaces traditional probabilistic feedback models with LLMs, using documents from the initial retrieval as generation context to expand the query, and significantly outperforms conventional methods.
HyDE~\citep{HyDE} prompts an LLM to generate a hypothetical relevant document and then encodes it as a pseudo query for similarity search.
GenRead~\citep{yu2023generate} builds on HyDE’s idea, showing that generating hypothetical documents can match or surpass pure retrieval-based pipelines.
More recently, CSQE~\citep{CSQE} combines sentences extracted from top-ranked corpus hits with LLM-generated expansions, outperforming fine-tuned neural query expanders on challenging queries.

Another line of work treats LLMs as \emph{semantic document indexers}.
These methods typically first leverage LLMs to analyze the documents through information extraction~\citep{doi:10.1177/0165551509360123, niklaus-etal-2018-survey}, including tasks such as text classification~\citep{weshclass, kargupta-etal-2023-megclass, teleclass} and entity extraction~\citep{jiao-etal-2023-instruct, qi-etal-2024-adelie, xu2025zeroshotopenschemaentitystructure}, and then perform fine-grained matching based on the extracted features.
TaxoIndex \citep{taxoindex} prompts LLMs to link documents with semantic entities drawn from the pre-defined set \citep{mag}.
MixGR~\citep{cai2024mixgr} decomposes queries into subqueries and documents into propositions, then computes and fuses multiple similarity scores using Reciprocal Rank Fusion \citep{rrf}. 
CoRank~\citep{tian2025corankllmbasedcompactreranking} extracts categories, topics, and keyphrases from documents, and integrates these features into compact document proxies to expand the range of listwise LLM reranking~\citep{sun-etal-2023-chatgpt}.
Most recently, SemRank~\citep{semrank} leverages LLMs to extract core concepts from scientific papers and queries, and employs these concepts for precise semantic matching.

Our method follows the semantic indexing paradigm, but differs by explicitly modeling structured entity-aspect pairs rather than treating entities or concepts as flat, independent units. 
This design enables more context-aware and fine-grained matching by capturing how scientific meaning arises from the interaction between entities and their associated aspects, addressing a key limitation of prior approaches that neglect such compositional structure.

\section{Methodology}
\begin{figure*}[ht!]
  \centering
  \includegraphics[width=1\linewidth]{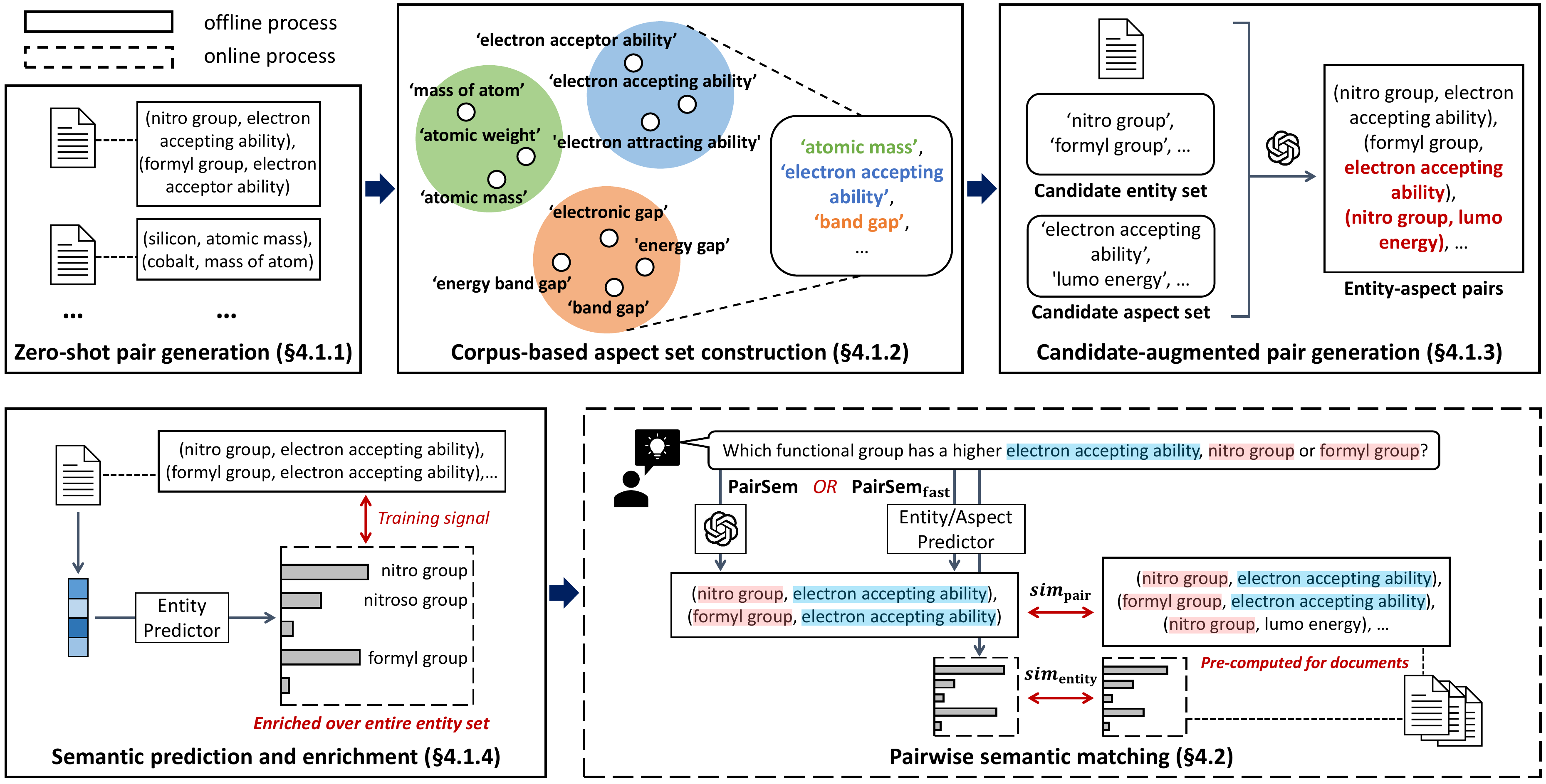}
  \vspace{-0.5cm}
  \caption{Overview of Pairwise Semantic Matching (\method) framework. The processes in \S\ref{Sec:4.1} are performed offline prior to query arrival, while only \S\ref{Sec:4.2} is executed at inference time.}
\label{fig:method}
\end{figure*}

We propose LLM-Guided \textbf{Pair}wise \textbf{Sem}antic Matching (\textbf{\method}) framework for a fine-grained mapping between scientific queries and documents.
We first describe the offline pre-processing stage (\S\ref{Sec:4.1}), where \method generates and enriches pairwise semantics for documents in corpus.
We then present the online inference stage (\S\ref{Sec:4.2}), where \method or \methodfast efficiently generates pairs for ad-hoc queries and leverages pairwise semantic matching to enhance scientific document retrieval.
The overview of \method is illustrated in Figure~\ref{fig:method}.

\subsection{LLM-Guided Pairwise Semantic Generation}\label{Sec:4.1}
To capture the fine-grained concepts of scientific documents, we propose to generate entity-aspect pairs from texts.
\textit{Entities} refer to objects that capture the core themes of scientific documents, such as chemical compounds like \texttt{"perfluorinated acid"}, or computer science models like \texttt{"gpt-oss-120b"}.
\textit{Aspects} are concepts associated with entities, such as the \texttt{"melting point"} of a chemical compound or the \texttt{"post-training"} of a language model.
In this paper, we aim to extract these entity-aspect pairs in an unsupervised manner, without relying on external data or human annotations.

\subsubsection{\textbf{Zero-shot pair generation}}
The straightforward approach to generating a set of entity-aspect pairs $\mathcal{P}^{\text{init}}_d$ from a document $d \in \mathcal{D}$ would be a zero-shot prompting to an LLM $\mathcal{M}_{\textup{LLM}}$:
\begin{equation}
    \mathcal{P}^{\text{init}}_d = \{ (e_i, a_i)\}_i \sim \mathcal{M}_{\textup{LLM}}(d  \mid  P_{\text{pair}}, P_{\text{format}}),
\label{eq:zeroshot}
\end{equation}
where $\sim$ denotes the generation process, and $P_{\text{pair}}$ represents the instruction for pair generation, as follows:
\begin{tcolorbox}[
    colback=gray!10,        
    colframe=black,          
    sharp corners,           
    boxrule=0.5pt,           
    left=1mm,                
    right=1mm,               
    top=0mm,                 
    bottom=0mm,              
    before skip=2mm,         
    after skip=2mm           
]
$P_{\text{pair}}=$ \texttt{"Given a scientific document, identify all scientific entities. Then, for each entity, find all associated aspects and generate (entity, aspect) pairs"}
\label{prompt:pair}
\end{tcolorbox}

\noindent Additionally, we leverage the ability of LLMs for the structured output \cite{liu2024westruct, yang2025structeval}, with a format prompt $P_{\text{format}}$.
\begin{tcolorbox}[
    colback=gray!10,        
    colframe=black,          
    sharp corners,           
    boxrule=0.5pt,           
    left=1mm,                
    right=1mm,               
    top=0mm,                 
    bottom=0mm,              
    before skip=2mm,         
    after skip=2mm           
]
$P_{\text{format}}=$ \texttt{"Output only (entity, aspect) pairs using}\\
\texttt{the following XML structure for each pair:\textcolor{red}{<pair>}}\\
\texttt{\textcolor{blue}{<entity>}entity name\textcolor{blue}{</entity>}\textcolor{orange}{<aspect>}aspect phrase}\\
\texttt{\textcolor{orange}{</aspect>}\textcolor{red}{</pair>}"}
\label{prompt:xml}
\end{tcolorbox}
\noindent In this manner, we can readily parse the output and construct a set of structured concepts from scientific documents.

While these semantic pairs are generated based on rich pretrained knowledge of LLMs, we observe that this approach has some challenges that need to be resolved:

\noindent \textbf{(C1) Synonym resolution}: We observe LLMs often output different surface names for the same concept. For example, for a chemical atom, \texttt{"atomic mass"} is represented by multiple synonyms, including \texttt{"atomic weight"} and \texttt{"mass of atom"}.

\noindent \textbf{(C2) Missing pairs}: We observe LLMs often generate an incomplete set of pairwise semantics, omitting relevant pairs that could be valuable matching signals for effective retrieval.

\noindent In the remaining part of this section, we present how \method builds corpus-based entity/aspect sets by merging synonyms (\S\ref{Sec:4.1.2}), and encourage LLMs to discover complete relevant pairs with candidate augmentation (\S\ref{Sec:4.1.3}).

\subsubsection{\textbf{Corpus-based entity/aspect set construction}}\label{Sec:4.1.2}
In this section, we aim to construct corpus-based entity/aspect sets, by merging synonyms and resolving \textbf{(C1)}.
We start by constructing initial entity and aspect sets ($\mathcal{E}^{\text{init}}$ and $\mathcal{A}^{\text{init}}$), by aggregating previously generated pairs across the corpus:
\begin{equation}
    \mathcal{E}^{\text{init}} = \bigcup_{d \in \mathcal{D}} \{ e \mid (e, a) \in \mathcal{P}^{\text{init}}_d \}, \,\,\,\mathcal{A}^{\text{init}} = \bigcup_{d \in \mathcal{D}} \{ a \mid (e, a) \in \mathcal{P}^{\text{init}}_d \}.
\end{equation}
We then gather semantically similar concepts with clustering.
Specifically, we embed all entities and aspects using the base retriever and cluster them hierarchically with agglomerative clustering \cite{ward1963hierarchical}.\footnote{We set the maximum size of clusters to 20.}
By doing so, we have two sets of clusters ($\mathcal{C}^{\mathcal{E}}$ and $\mathcal{C}^{\mathcal{A}}$), and each cluster $c \in \mathcal{C}^{*}$ contains semantically similar academic concepts.

However, we cannot naively merge those concepts in the same cluster to the same entity/aspect name.
For example, \texttt{"1-d structure"} and \texttt{"3-d structure"} are closely located in the embedding space, but they indeed represent distinct concepts that cannot be merged into a single aspect.
To tackle this, we exploit LLMs' strong entity comprehension ability \cite{llmclustering1, llmclustering2}. 
This allows us to readily integrate different surface names for the same concept, or distinguish distinct concepts under similar semantics.
Specifically, for each entity cluster $c \in \mathcal{C}^{\mathcal{E}}$, we prompt LLMs to detect a set of synonyms $s^*_j$ and merge them by generating a representative entity $e^*_j$.
\begin{equation}
    \{ (s^*_j, e^*_j)\}_j \sim \mathcal{M}_{\textup{LLM}}(c  \mid  P_{\text{cluster}}, P'_{\text{format}}),
\end{equation}
where $P_{\text{cluster}}$ is the instruction for the synonym merging.

\begin{tcolorbox}[
    colback=gray!10,        
    colframe=black,          
    sharp corners,           
    boxrule=0.5pt,           
    left=1mm,                
    right=1mm,               
    top=0mm,                 
    bottom=0mm,              
    before skip=2mm,         
    after skip=2mm           
]
$P_{\text{cluster}}=$ \texttt{"Given a list of scientific entities, find sets of synonyms that describe the same academic concept. Then, for each synonym set, generate a representative entity."}
\label{prompt:cluster}
\end{tcolorbox}

\noindent Similarly, a new format prompt $P'_{\text{format}}$ is provided as follows:
\begin{tcolorbox}[
    colback=gray!10,        
    colframe=black,          
    sharp corners,           
    boxrule=0.5pt,           
    left=1mm,                
    right=1mm,               
    top=0mm,                 
    bottom=0mm,              
    before skip=2mm,         
    after skip=2mm           
]
$P'_{\text{format}}=$ \texttt{"Output synonyms and representatives using}\\
\texttt{the following XML structure for each set:}\\
\texttt{\textcolor{red}{<set>}\textcolor{blue}{<entities>}entity1, entity2,...\textcolor{blue}{</entities>\textcolor{orange}{<rep>}}}\\
\texttt{representative entity\textcolor{orange}{</rep>}\textcolor{red}{</set>}}
\label{prompt:xml2}
\end{tcolorbox}

\noindent We also conduct this process for the aspects clusters $\mathcal{C}^{\mathcal{A}}$.
After parsing and aggregating the outputs, we get final sets of entities and aspects ($\mathcal{E}$ and $\mathcal{A}$).
Also, we devise $f^{\mathcal{E}}: \mathcal{E}^{\text{init}} \rightarrow \mathcal{E}$ and $f^{\mathcal{A}}: \mathcal{A}^{\text{init}} \rightarrow \mathcal{A}$, mappings from initial names to representatives.

\subsubsection{\textbf{Candidate-augmented pair generation}}\label{Sec:4.1.3}
We then use the constructed entity/aspect sets, to ground the pairwise semantic generation and ensure consistency in terminology across documents.
A straightforward approach would be to include the final sets $\mathcal{E}$ and $\mathcal{A}$ in the prompt and instruct $\mathcal{M}_{\text{LLM}}$ to generate pairs by selecting relevant entities and aspects from them.  
However, directly providing thousands of entities and aspects in the prompt is both inefficient and less effective in practice.

Instead, for each document $d \in \mathcal{D}$, we devise \textit{candidate} sets of entities and aspects ($\mathcal{E}^{\text{cand}}_d$ and $\mathcal{A}^{\text{cand}}_d$).
Specifically, we find three types of candidate entities as follows:
\begin{itemize}[leftmargin=10pt, itemsep=2pt]
    \item $\mathcal{E}^{\text{init}}_d = \{ f^{\mathcal{E}}(e) \mid (e,a) \in \mathcal{P}^{\text{init}}_d \}$ : Entities included in the initial pairs $\mathcal{P}^{\text{init}}_d$.
    \item $\mathcal{E}^{\text{lexical}}_d = \{ f^{\mathcal{E}}(e) \mid e \in d \}$ : Entities lexically included in the document $d$.
    \item $\mathcal{E}^{\text{pr}}_d =  \bigcup_{d' \in \mathcal{D}_d} \mathcal{E}^{\text{init}}_{d'}$: Pseudo-relevant entities included in the initial pairs of similar documents. $\mathcal{D}_d$ denotes the set of 10 documents nearest to $d$ in the embedding space.
\end{itemize}
We select the top-$M$ entities with the highest frequency in those three candidate sets and construct a final candidate entity set $\mathcal{E}^{\text{cand}}_d$.
In the same manner, we find a final candidate aspect set $\mathcal{A}^{\text{cand}}_d$.

Grounded on those candidate sets, we generate the final pairwise semantics $\mathcal{P}_d$ from a document $d \in \mathcal{D}$:
\begin{equation}
    \mathcal{P}_d = \{ (e_i, a_i)\}_i \sim \mathcal{M}_{\textup{LLM}}(d  \mid \mathcal{E}^{\text{cand}}_d, \mathcal{A}^{\text{cand}}_d, P'_{\text{pair}}, P_{\text{format}}).
\label{eq:cand}
\end{equation}
Here, we slightly modify the instruction $P_{\text{pair}}$ as follows:
\begin{tcolorbox}[
    colback=gray!10,        
    colframe=black,          
    sharp corners,           
    boxrule=0.5pt,           
    left=1mm,                
    right=1mm,               
    top=0mm,                 
    bottom=0mm,              
    before skip=2mm,         
    after skip=2mm           
]
$P'_{\text{pair}}=$\texttt{"Given a scientific document, find all relevant entities from \textcolor{red}{\{${\mathcal{E}^{\text{cand}}_d}$\}}. Then, for each entity, find all associated aspects from \textcolor{blue}{\{${\mathcal{A}^{\text{cand}}_d}$\}}, and generate relevant (entity, aspect) pairs"}
\label{prompt:candidate}
\end{tcolorbox}

\noindent This approach grounds pairwise semantic generation in our corpus-based sets of entities and aspects, thereby reducing hallucinations and enhancing consistency in expression.  
Furthermore, we observed that including candidates in the prompt prevents LLMs from missing the relevant pairs \textbf{(C2)}, thereby increasing the number of generated pairs (\S\ref{Sec:5.3.1}).

\subsubsection{\textbf{Semantic prediction and enrichment}}\label{Sec:4.1.4}
Based on the generated semantic pairs $\mathcal{P}_d$, we further enrich the information by estimating relevance between documents and entities/aspects.
By doing so, we aim to obtain soft probabilities of a document’s relevance to the entire entity/aspect sets, rather than binary indicators of their presence in the document’s semantic pairs.

To this end, we employ a lightweight entity predictor as follows:\footnote{Developing a novel architecture is not the main goal of this paper, and more sophisticated methods can be applied to our work.}
\begin{equation}
    \hat{y}_{d,e} = \hat{p}(e|d) = \sigma (\textbf{e}_e^\top f(\textbf{e}_d)),
\label{eq:entityr}
\end{equation}
where $\hat{y}_{d,e}$ is the estimated relevance between a document $d \in \mathcal{D}$ and an entity $e \in \mathcal{E}$.
$\sigma(\cdot)$ denotes the sigmoid function, and $f(\cdot)$ represents an $L$-layer MLP whose output dimension matches that of $\mathbf{e}_e$.
For the embeddings of entities ($\textbf{e}_e$) and documents ($\textbf{e}_d$), we employ the same base dense retriever used for retrieval.

A natural training signal for $\hat{y}_{d,e}$ would be a binary value $y_{d,e} \in \{0,1\}$, where $y_{d,e} = 1$ if $e \in \mathcal{E}_d$ and $0$ otherwise.\footnote{$\mathcal{E}_d = \{ e | (e,a) \in \mathcal{P}_d\}$}
However, not all entities contribute equally; entities have different levels of relevance to a given document.
For example, a popular entity like \texttt{"oxygen"} may be less distinctive than \texttt{"perfluorinated acid"} to represent academic documents.
To consider this, we adopt a distinctiveness metric inspired by \cite{taxocom}:
\begin{equation}
\text{DST}(d,e) = \frac{\exp\left( \text{BM25}(d,e) \right)}{1 + \sum_{d' \in \mathcal{D}_d} \exp\left( \text{BM25}(d',e) \right)},
\end{equation}
where $\mathcal{D}_d$ denotes the set of 10 documents nearest to $d$ in the embedding space.
We then normalize the distinctiveness scores to produce a soft target value $y_{d,e} \in [0,1]$:
\begin{equation}
y_{d,e} = \frac{\text{DST}(d,e)}{\max_{e' \in \mathcal{E}} \text{DST}(d,e') }.
\label{eq:softlabel}
\end{equation}
Finally, we train the predictor by using a binary cross-entropy loss:
\begin{equation}
\mathcal{L}_{\text{e}} = - \sum_{d \in \mathcal{D}} \Big( \sum_{e \in \mathcal{E}_d} y_{d,e} \log \hat{y}_{d,e} +  \sum_{e \notin \mathcal{E}_d} \log (1-\hat{y}_{d,e})  \Big).
\label{eq:BCE}
\end{equation}

Similarly, we devise an aspect relevance estimator as follows:
\begin{equation}
\hat{y}_{d,a\mid e} = \hat{p}(a|d,e) = \sigma (\textbf{e}_a^{\top} \, g(\textbf{e}_d:\textbf{e}_e)),
\label{eq:aspectr}
\end{equation}
where $:$ is the concatenation, and $g(\cdot)$ is an $L$-layer MLP whose output dimension matches that of $\mathbf{e}_a$.
Since each aspect is always associated with an entity, we estimate the relevance between a document $d$ and an aspect $a$, conditioned on the entity $e$.
The aspect predictor is trained with the following loss:
\begin{equation}
\mathcal{L}_{a} = - \sum_{d \in \mathcal{D}} \sum_{e \in \mathcal{E}_d} \sum_{a \in \mathcal{A}} \Big( y_{d,a \mid e} \log \hat{y}_{d,a \mid e} +  (1-y_{d,a\mid e})\log (1-\hat{y}_{d,a \mid e})  \Big),
\label{eq:BCE2}
\end{equation}
where $y_{d,a \mid e} = 1$ if $(e,a)\in\mathcal{P}_d$, and $0$ otherwise.
By using the learned predictors, we estimate entity relevance $\hat{y}_{d,e}$ and aspect relevance $\hat{y}_{d,a\mid e}$ that provide soft probabilities of a document’s relevance to the entire entity/aspect sets.

\subsection{Pairwise Semantic Matching at Inference}\label{Sec:4.2}
So far, we have generated document-side semantic pairs ($\mathcal{P}_d$) along with the estimated entity and aspect relevances ($\hat{y}_{d,e}$ and $\hat{y}_{d,a \mid e}$).  
Next, we describe how query-side semantic pairs are generated and how semantic information from both queries and documents is utilized during inference.  

\subsubsection{\textbf{Online query semantic generation}}\label{sec:4.2.1}
At inference, we need to generate pairwise semantics for the ad-hoc queries.
As the simplest choice, we can apply the same process used for documents (\S\ref{Sec:4.1.3}).
We refer to this approach as \textbf{PairSem}.

However, \method requires additional LLM inferences at test time to generate query-side pairwise semantics, which can be impractical under strict time constraints.  
To address this, we replace the LLM-based generation with the learned entity and aspect predictors ($\hat{y}_{q,e}$ and $\hat{y}_{q,a \mid e}$)\footnote{Obtained by substituting $\textbf{e}_d$ with $\textbf{e}_q$ in Eq.\ref{eq:entityr} and Eq.\ref{eq:aspectr}.}, and generate query pairs through the following steps:
\begin{enumerate}[leftmargin=15pt, itemsep=2pt]
    \item We first select top-$N^e$ entities with estimated relevance $\hat{y}_{q,e}$.
    \item For each selected entity $e$, we select top-$N^a$ aspects with estimated relevance $\hat{y}_{q,a \mid e}$.
\end{enumerate}
By doing so, we construct $\mathcal{P}_q$, a set of semantic pairs for a query $q$, without using an LLM at inference time.
In this work, we set $N^e$=10 and $N^a$=5.
We refer this approach as \textbf{PairSem$_{\text{fast}}$}.

\begin{table*}[t]
\caption{Retrieval performance of various enhancement methods. * indicates statistical significance (paired t-test with $p<0.05$) compared to the best competitor. We also conducted the t-test compared to Base, and \textit{all} results are statistically significant.}
\centering
\renewcommand{\arraystretch}{0.9}
\resizebox{\linewidth}{!}{
\begin{tabular}{cl cccc cccc cccc}\toprule
& \multicolumn{1}{c}{\multirow{2}{*}{Method}} & \multicolumn{4}{c}{ChemLit-QA} & \multicolumn{4}{c}{SciFact} & \multicolumn{4}{c}{LitSearch} \\
\cmidrule(lr){3-6} \cmidrule(lr){7-10} \cmidrule(lr){11-14}
& & N@10 & N@20 & R@20 & R@50 & N@10 & N@20 & R@20 & R@50 & N@10 & N@20 & R@20 & R@50 \\
\midrule \midrule

\parbox[t]{2mm}{\multirow{10}{*}{\rotatebox[origin=c]{90}{SPECTER2}}} 
& Base & 0.5279 & 0.5968 & 0.6923 & 0.8132 & 0.6616 & 0.6701 & 0.8382 & 0.9090 & 0.3335 & 0.3520 & 0.5567 & 0.6547 \\
\cmidrule(lr){2-14} 
& BERT-QE \cite{zheng2020bertqe} & 0.5263 & 0.6089 & 0.6942 & 0.8149 & 0.6594 & 0.6724 & 0.8510 & 0.9119 & 0.3276 & 0.3487 & 0.5454 & 0.6486 \\
& LADR \cite{LADR} & 0.5311 & 0.6103 & 0.7064 & 0.8212 & 0.6656 & 0.6834 & 0.8580 & 0.9203 & 0.3408 & 0.3671 & 0.5723 & 0.6641 \\
& ToTER \cite{toter} & 0.5488 & 0.6114 & 0.7190 & 0.8310 & 0.6642 & 0.6815 & 0.8581 & 0.9147 & 0.3401 & 0.3662 & 0.5719 & 0.6655 \\
& \cellcolor{gray!20}\textbf{PairSem$_{\text{fast}}$} & 
\cellcolor{gray!20}\textbf{0.5650}* & \cellcolor{gray!20}\textbf{0.6374}* & \cellcolor{gray!20}\textbf{0.7468}* & \cellcolor{gray!20}\textbf{0.8625}* &
\cellcolor{gray!20}\textbf{0.6846}* & \cellcolor{gray!20}\textbf{0.7010}* & \cellcolor{gray!20}\textbf{0.8844}* & \cellcolor{gray!20}\textbf{0.9330}* &
\cellcolor{gray!20}\textbf{0.3579}* & \cellcolor{gray!20}\textbf{0.3808}* & \cellcolor{gray!20}\textbf{0.5861}* & \cellcolor{gray!20}\textbf{0.6869}* \\
\cmidrule(lr){2-14} 
& HyDE \cite{HyDE}& 0.5534 & 0.6081 & 0.7289 & 0.8319 & 0.6776 & 0.6842 & 0.8556 & 0.9107 & 0.3548 & 0.3767 & 0.6025 & 0.6954 \\
& GRF \cite{grf}& 0.5581 & 0.6162 & 0.7330 & 0.8439 & 0.6894 & 0.6961 & 0.8677 & 0.9221 & 0.3512 & 0.3729 & 0.5921 & 0.6913 \\
& CSQE \cite{CSQE} & 0.5608 & 0.6121 & 0.7303 & 0.8322 & 0.6713 & 0.6881 & 0.8607 & 0.9164 & 0.3496 & 0.3718 & 0.5868 & 0.6901 \\
& SemRank \cite{semrank} & 0.5620 & 0.6318 & 0.7479 & 0.8459 & 0.6910 & 0.7001 & 0.8682 & 0.9250 & 0.3657 & 0.3804 & 0.6324 & 0.7017 \\
& \cellcolor{gray!20}\textbf{PairSem} & 
\cellcolor{gray!20}\textbf{0.5903}* & \cellcolor{gray!20}\textbf{0.6617}* & \cellcolor{gray!20}\textbf{0.7623}* & \cellcolor{gray!20}\textbf{0.8687}* &
\cellcolor{gray!20}\textbf{0.7100}* & \cellcolor{gray!20}\textbf{0.7282}* & \cellcolor{gray!20}\textbf{0.9056}* & \cellcolor{gray!20}\textbf{0.9403}* &
\cellcolor{gray!20}\textbf{0.3845}* & \cellcolor{gray!20}\textbf{0.4037}* & \cellcolor{gray!20}\textbf{0.6474}* & \cellcolor{gray!20}\textbf{0.7229}* \\

\midrule
\parbox[t]{2mm}{\multirow{10}{*}{\rotatebox[origin=c]{90}{Contriever-MS}}} 
& Base & 0.6279 & 0.6891 & 0.7431 & 0.8344 & 0.6768 & 0.6898 & 0.8619 & 0.9120 & 0.3694 & 0.3784 & 0.5753 & 0.6627 \\
\cmidrule(lr){2-14} 
& BERT-QE \cite{zheng2020bertqe} & 0.6131 & 0.6877 & 0.7415 & 0.8349 & 0.6702 & 0.6839 & 0.8576 & 0.9114 & 0.3681 & 0.3799 & 0.5768 & 0.6631 \\
& LADR \cite{LADR} & 0.6334 & 0.6921 & 0.7490 & 0.8411 & 0.6783 & 0.6921 & 0.8661 & 0.9205 & 0.3755 & 0.3923 & 0.5906 & 0.6649 \\
& ToTER \cite{toter} & 0.6319 & 0.6916 & 0.7475 & 0.8408 & 0.6770 & 0.6904 & 0.8638 & 0.9137 & 0.3805 & 0.3980 & 0.5964 & 0.6678 \\
& \cellcolor{gray!20}\textbf{PairSem$_{\text{fast}}$} & 
\cellcolor{gray!20}\textbf{0.6438}* & \cellcolor{gray!20}\textbf{0.7086}* & \cellcolor{gray!20}\textbf{0.7641}* & \cellcolor{gray!20}\textbf{0.8552}* &
\cellcolor{gray!20}\textbf{0.6922}* & \cellcolor{gray!20}\textbf{0.7060}* & \cellcolor{gray!20}\textbf{0.8824}* & \cellcolor{gray!20}\textbf{0.9377}* &
\cellcolor{gray!20}\textbf{0.3990}* & \cellcolor{gray!20}\textbf{0.4189}* & \cellcolor{gray!20}\textbf{0.6224}* & \cellcolor{gray!20}\textbf{0.6889}* \\
\cmidrule(lr){2-14} 
& HyDE \cite{HyDE} & 0.6304 & 0.6987 & 0.7481 & 0.8432 & 0.6818 & 0.6942 & 0.8705 & 0.9154 & 0.3714 & 0.3826 & 0.5801 & 0.6718 \\
& GRF \cite{grf} & 0.6342 & 0.6976 & 0.7558 & 0.8457 & 0.6974 & 0.7011 & 0.8731 & 0.9201 & 0.3846 & 0.3941 & 0.5986 & 0.6811 \\
& CSQE \cite{CSQE} & 0.6217 & 0.6837 & 0.7441 & 0.8365 & 0.6834 & 0.6981 & 0.8636 & 0.9141 & 0.3853 & 0.3935 & 0.5864 & 0.6723 \\
& SemRank \cite{semrank} & 0.6335 & 0.7014 & 0.7646 & 0.8424 & 0.6931 & 0.7096 & 0.8655 & 0.9187 & 0.3854 & 0.4067 & 0.6057 & 0.6807 \\
& \cellcolor{gray!20}\textbf{PairSem} & 
\cellcolor{gray!20}\textbf{0.6501}* & \cellcolor{gray!20}\textbf{0.7121}* & \cellcolor{gray!20}\textbf{0.7756}* & \cellcolor{gray!20}\textbf{0.8612}* &
\cellcolor{gray!20}\textbf{0.7112}* & \cellcolor{gray!20}\textbf{0.7239}* & \cellcolor{gray!20}\textbf{0.8859}* & \cellcolor{gray!20}\textbf{0.9427}* &
\cellcolor{gray!20}\textbf{0.4139}* & \cellcolor{gray!20}\textbf{0.4314}* & \cellcolor{gray!20}\textbf{0.6241}* & \cellcolor{gray!20}\textbf{0.6973}* \\

\midrule
\parbox[t]{2mm}{\multirow{10}{*}{\rotatebox[origin=c]{90}{Qwen3-Embedding-8B}}} 
& Base & 0.6795 & 0.7491 & 0.8384 & 0.9149 & 0.7695 & 0.7759 & 0.9300 & 0.9567 & 0.5802 & 0.5982 & 0.7916 & 0.8400 \\
\cmidrule(lr){2-14} 
& BERT-QE \cite{zheng2020bertqe} & 0.6613 & 0.7338 & 0.8227 & 0.9103 & 0.7538 & 0.7681 & 0.9257 & 0.9495 & 0.5811 & 0.5995 & 0.7939 & 0.8445 \\
& LADR \cite{LADR} & 0.6801 & 0.7536 & 0.8411 & 0.9170 & 0.7717 & 0.7812 & 0.9373 & 0.9603 & 0.5906 & 0.6042 & 0.7989 & 0.8532 \\
& ToTER \cite{toter} & 0.6823 & 0.7560 & 0.8476 & 0.9205 & 0.7809 & 0.7875 & 0.9417 & 0.9583 & 0.5948 & 0.6078 & 0.8022 & 0.8577 \\
&\cellcolor{gray!20} \textbf{PairSem$_{\text{fast}}$} & 
\cellcolor{gray!20}\textbf{0.6959}* & \cellcolor{gray!20}\textbf{0.7663}* & \cellcolor{gray!20}\textbf{0.8599}* & \cellcolor{gray!20}\textbf{0.9297} &
\cellcolor{gray!20}\textbf{0.7900}* & \cellcolor{gray!20}\textbf{0.7959} & \cellcolor{gray!20}\textbf{0.9467} & \cellcolor{gray!20}\textbf{0.9667}* &
\cellcolor{gray!20}\textbf{0.6125}* & \cellcolor{gray!20}\textbf{0.6263}* & \cellcolor{gray!20}\textbf{0.8103} & \cellcolor{gray!20}\textbf{0.8634} \\
\cmidrule(lr){2-14} 
& HyDE \cite{HyDE} & 0.6720 & 0.7471 & 0.8367 & 0.9201 & 0.7584 & 0.7647 & 0.9261 & 0.9445 & 0.5728 & 0.5961 & 0.7814 & 0.8338 \\
& GRF \cite{grf} & 0.6813 & 0.7502 & 0.8438 & 0.9161 & 0.7728 & 0.7816 & 0.9354 & 0.9526 & 0.5854 & 0.6009 & 0.7946 & 0.8501 \\
& CSQE \cite{CSQE} & 0.6735 & 0.7428 & 0.8341 & 0.9113 & 0.7615 & 0.7728 & 0.9312 & 0.9570 & 0.5791 & 0.5977 & 0.7897 & 0.8395 \\
& SemRank \cite{semrank} & 0.6800 & 0.7494 & 0.8446 & 0.9171 & 0.7824 & 0.7884 & 0.9407 & 0.9601 & 0.5866 & 0.6018 & 0.7935 & 0.8518 \\
& \cellcolor{gray!20}\textbf{PairSem} & 
\cellcolor{gray!20}\textbf{0.6978}* & \cellcolor{gray!20}\textbf{0.7661}* & \cellcolor{gray!20}\textbf{0.8560}* & \cellcolor{gray!20}\textbf{0.9280}* &
\cellcolor{gray!20}\textbf{0.7906}* & \cellcolor{gray!20}\textbf{0.7970}* & \cellcolor{gray!20}\textbf{0.9433} & \cellcolor{gray!20}\textbf{0.9700}* &
\cellcolor{gray!20}\textbf{0.6113}* & \cellcolor{gray!20}\textbf{0.6273}* & \cellcolor{gray!20}\textbf{0.8184}* & \cellcolor{gray!20}\textbf{0.8751}* \\

\bottomrule
\end{tabular}}
\label{tab:main}
\end{table*}

\subsubsection{\textbf{Enhancing retrieval with pairwise semantic matching}}\label{Sec:4.2.2}
Finally, based on the semantic pairs of a query ($\mathcal{P}_q$) and a document ($\mathcal{P}_d$), we define a pairwise semantic matching score as follows:
\begin{equation}
sim_{\text{pair}}(q,d) = \frac{1}{|\mathcal{P}_q|} \sum_{(e,a) \in \mathcal{P}_q} \max_{(e',a') \in \mathcal{P}_d} \mathbbm{1}[e=e'] \cdot sim_{\text{base}}(a,a'),
\label{eq:pair}
\end{equation}
where $\mathbbm{1}[\cdot]$ denotes the indicator function, which returns 1 when its argument evaluates to true and 0 otherwise.
$sim_{\text{base}}(a,a')$ denotes the aspect-aspect similarity, measured by the base retriever (i.e., embedding similarity between them).
Our pairwise semantic matching score measures the similarity between semantic pairs of a query and a document, by identifying the most similar aspect for each matched entity.

Additionally, we utilize the learned document-entity relevance ($\hat{y}_{d,e}$) as well as the inferred query-entity relevance ($\hat{y}_{q,e}$).
The enriched entity relevance provides a soft distribution over the entire entity set, complementing $sim_{\text{pair}}(q,d)$ that accounts only for the generated entities.
Specifically, we measure the KL-divergence between two entity relevance distributions:
\begin{equation}
sim_{\text{entity}}(q,d) = - \text{D}_{\text{KL}} (\hat{\textbf{y}}_{q} \,\, || \,\, \hat{\textbf{y}}_{d}),
\label{eq:entity}
\end{equation}
where $\hat{\textbf{y}}_{q}$ = $\{ \hat{y}_{q,e}\}_{e\in\mathcal{E}}$ and $\hat{\textbf{y}}_{d}$ = $\{ \hat{y}_{d,e}\}_{e\in\mathcal{E}}$.
It can be readily computed as the inner product between $\hat{\textbf{y}}_{q}$ and the logarithm of $\hat{\textbf{y}}_{d}$.

Finally, we combine the proposed similarity scores with the base retriever's one:
\begin{equation}
sim(q,d) = h(sim_{\text{base}}(q,d)) + \frac{h(sim_{\text{pair}}(q,d)) + h(sim_{\text{entity}}(q,d))}{2},
\label{eq:final}
\end{equation}
where $h(\cdot)$ is a normalization function (e.g., z-score or reciprocal rank\footnote{$h(sim_{*}(q,d))=1/ (1+rank(sim_{*}(q,d)))$.} \cite{rrf}) used to adjust the different score scales.
We adopt the latter one, as it yields slightly better performance in our experiment.

\section{Experiment}
\subsection{Experimental Setup}
We briefly summarize the experiment setup due to limited space.
Please refer to Appendix~\ref{sec:appenexp} for the experimental details.

\subsubsection{\textbf{Datasets}}
We select three datasets across different academic domains: \textbf{ChemLit-QA} (chemistry) \cite{wellawatte2024chemlitqa}, \textbf{SciFact} (biomedical) \cite{scifact}, and \textbf{LitSearch} (computer science) \cite{ajith2024litsearch}.
ChemLit-QA provides triplets of (query, relevant documents, answer), from which we use only the query-document pairs.
SciFact and LitSearch provide test query sets along with human-annotated relevance labels.
Notably, we do not use any training queries or entity/aspect information associated with the documents.

\subsubsection{\textbf{Evaluation metrics}}
We use NDCG@$K$ (N@$K$) \cite{jarvelin2002cumulated} and Recall@$K$ (R@$K$) as our evaluation metrics.
NDCG emphasizes the ranking precision among top candidates, while Recall captures retrieval coverage over a broader range.
Following previous studies \cite{semrank, xu2024bmretriever}, we report NDCG at $K \in \{10, 20\}$ and Recall at $K \in \{20, 50\}$.

\subsubsection{\textbf{Base retrievers}}
We employ two representative dense retrievers: \textbf{SPECTER2} \cite{specterv2} and \textbf{Contriever-MS} \cite{contriever}.
SPECTER2 is trained on scientific metadata (e.g., authors, citations), while Contriever-MS is fine-tuned on the MS MARCO dataset \cite{ms-marco}, which contains (q,d) pairs from general domains.
Additionally, we adopt \textbf{Qwen3-Embedding-8B} \cite{yang2025qwen3}, an LLM-based embedding model that achieves state-of-the-art performance across diverse retrieval tasks.
All base retrievers are sourced from their official Hugging Face repositories \cite{huggingface}.

\begin{figure}[t]
  \centering
  \includegraphics[width=1\linewidth]{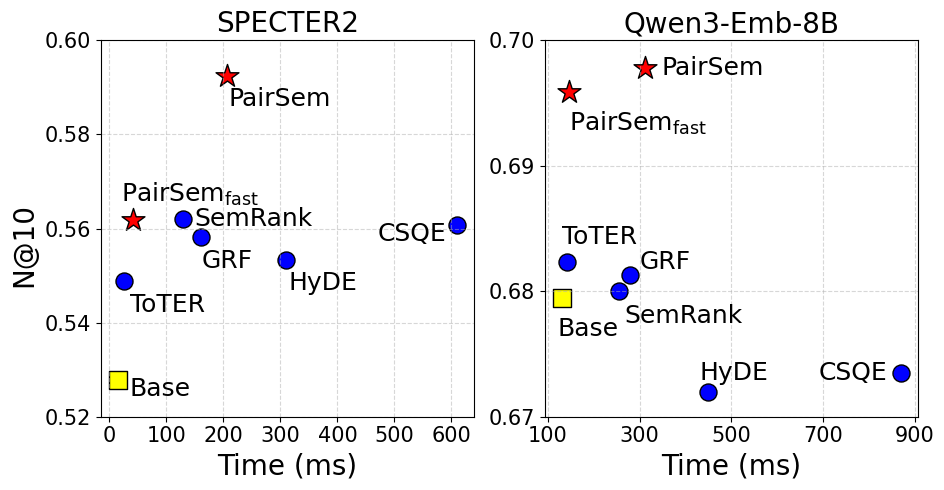}
  \vspace{-0.5cm}
  \caption{Time-accuracy trade-off on ChemLit-QA.}
  \vspace{-0.3cm}
\label{fig:time}
\end{figure}

\subsubsection{\textbf{Compared methods}}\label{Sec:5.1.4}
We compare \textbf{$\text{\method}_{\text{fast}}$} with methods that enhance base retrievers by leveraging corpus knowledge:
\begin{itemize}[leftmargin=10pt, itemsep=2pt]
    \item \textbf{BERT-QE} \cite{zheng2020bertqe} extracts informative \underline{text segments} from initial top-ranked documents, for augmenting queries.
    \item \textbf{LADR} \cite{LADR} employs a \underline{hybrid strategy} that first applies a lexical retriever to reduce the candidate pool for the base retriever.
    \item \textbf{ToTER} \cite{toter} enriches retrieval by linking documents and queries to academic concepts derived from an external \underline{topical taxonomy}.
\end{itemize}
We also compare \textbf{\method} with methods that utilize LLMs to generate relevant semantics:
\begin{itemize}[leftmargin=10pt, itemsep=2pt]
    \item \textbf{HyDE} \cite{HyDE} prompts an LLM to produce a \underline{hypothetical document} relevant to the query, which is then used as a pseudo-query.
    \item \textbf{GRF} \cite{grf} prompts an LLM to generate semantic features (e.g., \underline{keywords}, \underline{concepts}) to enhance the retrieval.
    \item \textbf{CSQE} \cite{CSQE} prompts an LLM to extract \underline{salient sentences} from top-ranked documents and expands queries with them.
    \item \textbf{SemRank} \cite{semrank} prompts an LLM to associate queries and documents with relevant \underline{topics and phrases} drawn from an external concept set.
\end{itemize}
For fairness, we use \texttt{GPT-4.1-mini} \cite{achiam2023gpt4} as the backbone LLM for \method (ours) and all LLM-based baselines.

\subsubsection{\textbf{Implementation details}}
In \S\ref{Sec:4.1.3}, we set $M$, the number of candidate entities/aspects, to 50.
In \S\ref{Sec:4.1.4}, we set $L$, the number of layers in the entity/aspect predictors, to 2 for SPECTER2 and Contriever-MS, and 5 for Qwen3-Embedding-8B, reflecting their embedding dimensions.
All baselines are implemented using publicly available author code, with configurations and hyperparameters strictly following the documentation.
All experiments are conducted on a single NVIDIA A100 80GB GPU paired with an AMD EPYC™ 7513 2.60 GHz CPU.

\begin{table}[t]
\caption{Statistics on pair generation for corpus (\S\ref{Sec:4.1}).}
\label{tab:statpair}
\vspace{-0.3cm}
 \resizebox{1\columnwidth}{!}{
  \begin{tabular}{cccc}
    \toprule
    Procedure & Output & Time & Cost \\
    \midrule
    Zero-shot pair generation & 13.55 pairs / doc. & 84s & \$0.85  \\
    Entity set construction & 5,032 $\rightarrow$ 3,352 entities & 13s & \$0.20  \\
    Aspect set construction & 4,649 $\rightarrow$ 2,813 aspects & 11s & \$0.18  \\
        Candidate-augmented generation & 21.86 pairs / doc. & 113s & \$1.07  \\
    \bottomrule
  \end{tabular}}
  \vspace{-0.3cm}
\end{table}

\begin{figure}[t]
  \centering
  \includegraphics[width=1\linewidth]{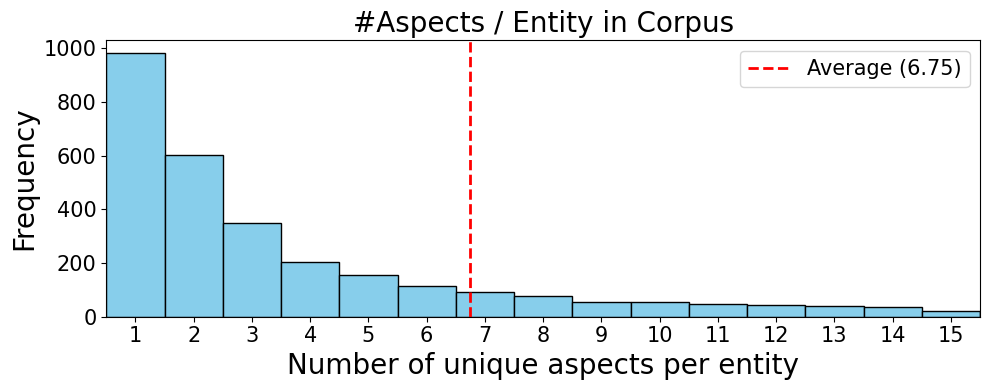}
  \vspace{-0.5cm}
  \caption{Number of unique aspects per entity in corpus.}
  \vspace{-0.3cm}
\label{fig:statpair}
\end{figure}

\subsection{Performance Comparison}
\subsubsection{\textbf{Retrieval performance}}
Table~\ref{tab:main} shows the retrieval performance of various enhancement methods.
For LLM-based methods, we conduct three independent runs to account for the stochastic generation of LLMs.
Overall, \method consistently outperforms all baselines across the datasets.
Specifically, \method improves the retrieval performance of the base retriever by up to 16.29\% (R@20 with SPECTER2 on LitSearch), without using any relevance labels or entity annotations.
Compared to SemRank, the strongest baseline, \method achieves gains of up to 7.39\% (N@10 with Contriever-MS on LitSearch).
Notably, \methodfast, which does not use LLMs to generate pairs for queries, shows comparable or even slightly better performance (0.7219 vs 0.7154, on average) than SemRank, which utilizes LLMs at inference.
This result demonstrates the effectiveness of our pairwise semantic matching approach and efficient inference with \methodfast.

\subsubsection{\textbf{Time-accuracy trade-off}}
Figure~\ref{fig:time} illustrates the time vs accuracy trade-off of the proposed methods and baseline approaches on the ChemLit-QA dataset.  
The evaluation uses two base retrievers: SPECTER2 and Qwen3-Embedding-8B.  
The x-axis represents the average query processing time, and the y-axis shows the retrieval performance (NDCG@10).
All LLM-based baselines employ the same backbone LLM for a fair comparison, as described in \S\ref{Sec:5.1.4}.  
Notably, \method requires only a single LLM call at inference, while \methodfast eliminates LLM usage in query processing.
Our structured pair output with XML-based prompting generates fewer tokens than methods producing multiple sentences for semantic expansion (e.g., HyDE, CSQE).
Consequently, both \method and \methodfast achieve the most favorable trade-off between efficiency and retrieval effectiveness.

\begin{table}[t]
\centering
\caption{Ablation study on PairSem.}
\label{tab:ablation}
\vspace{-0.3cm}
 \resizebox{1\columnwidth}{!}{
\begin{tabular}{lcccc}
\toprule
Method & N@10 & N@20 & R@20 & R@50 \\
\toprule
SPECTER2 (base)                & 0.5279 & 0.5968 & 0.6923 & 0.8132 \\
\midrule
\multicolumn{5}{l}{\textit{Generation} (\S\ref{Sec:4.1})} \\
\quad No synonym merging               & 0.5623 & 0.6335 & 0.7376 & 0.8418 \\
\quad No candidate & 0.5324 & 0.6083 & 0.7176 & 0.8363 \\
\quad No distinctiveness       & 0.5807 & 0.6515 & 0.7535 & 0.8579 \\
\midrule
\multicolumn{5}{l}{\textit{Matching} (\S\ref{Sec:4.2})} \\
\quad Flatten pairs                 & 0.5706 & 0.6431 & 0.7431 & 0.8468 \\
\quad No $sim_{\text{pair}}$             & 0.5412 & 0.6117 & 0.7147 & 0.8301 \\
\quad No $sim_{\text{entity}}$           & 0.5698 & 0.6436 & 0.7496 & 0.8492 \\
\midrule
\textbf{PairSem}                  & \textbf{0.5903} & \textbf{0.6617} & \textbf{0.7623} & \textbf{0.8687} \\
\bottomrule
\end{tabular}}
\vspace{-0.3cm}
\end{table}

\subsection{Study of \method}
Supplementary results (e.g., evaluation of entity/aspect predictors and hyperparameter study) are presented in Appendix~\ref{sec:appenresult}.

\subsubsection{\textbf{Analysis on pair generation}} \label{Sec:5.3.1}
Table~\ref{tab:statpair} reports statistics on pair generation for the ChemLit-QA corpus. 
For each step in \S\ref{Sec:4.1}, we present the time required for LLM generation (Time), the corresponding API costs (Cost), and the output statistics (Output).
From 1,528 documents, \method constructs sets of 3,352 entities (merged from 5,032) and 2,813 aspects (merged from 4,649), and generates 21.86 pairs per document on average.
It is worth noting that the average number of generated pairs is substantially increased when the pair generation is augmented with candidate entities and aspects (13.55 $\rightarrow$ 21.86).
Figure~\ref{fig:statpair} further presents the histogram of the number of unique aspects linked to each entity in the corpus.
Among the 3,352 entities, more than half are associated with more than two aspects, and the average number of unique aspects per entity is 6.75.
This supports our motivation for generating pairwise semantics rather than relying on flat entity sets.

\subsubsection{\textbf{Ablation study}}
We conduct ablation studies to demonstrate the effectiveness of each component of \method.
Specifically, we devise the following ablated variants:
\begin{itemize}[leftmargin=10pt, itemsep=2pt]
    \item \textbf{No synonym merging} (\S\ref{Sec:4.1.2}): excludes the synonym merging when constructing entity/aspect sets.
    \item \textbf{No candidate} (\S\ref{Sec:4.1.3}): excludes the candidate entities and aspects when generating pairs (i.e., zero-shot pair generation).
    \item \textbf{No distinctiveness} (\S\ref{Sec:4.1.4}): excludes the distinctiveness-aware soft labels (Eq.\ref{eq:softlabel}) for the entity predictor.
    \item \textbf{Flatten pairs} (\S\ref{Sec:4.2.2}): deconstructs the generated pairs into a single flat set of entities and aspects.
    \item \textbf{No $sim_{\text{pair}}$} (\S\ref{Sec:4.2.2}): excludes $sim_{\text{pair}}$ from final score (Eq.\ref{eq:final}).
    \item \textbf{No $sim_{\text{entity}}$} (\S\ref{Sec:4.2.2}): excludes $sim_{\text{entity}}$ from final score (Eq.\ref{eq:final}).
\end{itemize}
\noindent Table~\ref{tab:ablation} shows the result of ablation studies on ChemLit-QA dataset with SPECTER2 as the base retriever.
First, for Generation (\S\ref{Sec:4.1}), we observe that either excluding set construction with synonym merging or candidate-augmented pair generation degrades retrieval performance, as they ensure the generated pairs are grounded on the corpus and comprehensively cover the relevant concepts.
Second, for Matching (\S\ref{Sec:4.2}), flattening the generated pairs or excluding the pairwise matching score results in severe performance degradation.
This indicates that our pairwise semantics captures more fine-grained scientific concepts by considering interplay between entities and their associated aspects.

\subsubsection{\textbf{Results with domain-specific retrievers}}
While our main objective is to enhance general retrievers for specific target domains, we also demonstrate that \method can improve domain-specific retrievers without relying on human annotations.
We select two recent biomedical retrievers, \textbf{MedCPT} \cite{jin2023medcpt} and \textbf{BMRetriever-1B} \cite{xu2024bmretriever}. 
MedCPT is trained on large-scale search logs and papers from PubMed, and BMRetriever-1B is trained on biomedical corpora such as bioRxiv\footnote{https://www.biorxiv.org/} and multiple medical QA datasets \cite{medicalqa, herlihy2021mednli, medmcqa}. 

Table~\ref{tab:biomedical} shows that \method improves their performance even though both retrievers are already fine-tuned on domain-specific data.
In biomedical contexts, capturing entities (e.g., \texttt{"homocysteine"}) and linking them with specific aspects (e.g., \texttt{"blood level"}) is critical.
Despite domain-specific fine-tuning, it still remains difficult for holistic document-level embeddings to capture such detailed and structured semantics \cite{toter, shavarani2025entity}.
By generating pairwise semantics and explicitly leveraging them for matching, \method successfully enhances the effectiveness of these domain-specific retrievers.

\begin{table}[t]
\caption{Results with biomedical retrievers on SciFact.}
\label{tab:biomedical}
\vspace{-0.3cm}
 \resizebox{0.85\columnwidth}{!}{
  \begin{tabular}{l cccc}
    \toprule
    Model & N@10 & N@20 & N@50 & N@100  \\
    \toprule
    MedCPT & 0.7246 & 0.7417 & 0.7507 & 0.7531  \\
    \quad \textbf{+ PairSem} & \textbf{0.7473} & \textbf{0.7619} & \textbf{0.7665} & \textbf{0.7690} \\
    \midrule
    BMRetriever-1B & 0.7573 & 0.7671 & 0.7718 & 0.7742  \\
    \quad \textbf{+ PairSem} & \textbf{0.7673} & \textbf{0.7794} & \textbf{0.7890} & \textbf{0.7909} \\
    \bottomrule
  \end{tabular}}
  \vspace{-0.3cm}
\end{table}


\section{Conclusion}
We propose \method, a novel framework for fine-grained semantic matching between scientific queries and documents.
\method captures the interplay between entities and their associated aspects through pairwise semantics, enhancing embedding-based retrieval.
We ensure consistent terminology via corpus-level synonym merging, and augment pair generation with candidate entities and aspects to prevent omission of relevant pairs.
Our extensive experiments across three datasets and retrievers validate the effectiveness of \method over flat, independent semantic approaches.
Future work will explore integrating pairwise semantics into retriever training for further improvement.

\begin{acks}
This work was supported by Samsung Research Funding \& Incubation Center of Samsung Electronics (SRFC-IT2402-05, South Korea), by MSIT (No. RS-2024-00335873, South Korea), and by Molecule Maker Lab Institute: An AI Research Institutes program supported by NSF (No. 2019897, United States).
\end{acks}

\bibliographystyle{ACM-Reference-Format}
\bibliography{reference}

\clearpage

\appendix
\section{Appendix}
\subsection{Experimental Details}\label{sec:appenexp}
\subsubsection{\textbf{Data \& model sources}}  
We use three datasets from their official repositories:  
\begin{itemize}[leftmargin=10pt, itemsep=2pt]  
    \item \textbf{ChemLit-QA:} https://github.com/geemi725/ChemLit-QA  
    \item \textbf{SciFact:} https://huggingface.co/datasets/allenai/scifact  
    \item \textbf{LitSearch:} https://huggingface.co/datasets/princeton-nlp/LitSearch  
\end{itemize}  
Table~\ref{tab:dataset} presents the data statistics.

\begin{table}[h!]
\caption{Data statistics.}
\label{tab:dataset}
\vspace{-0.3cm}
 \resizebox{0.8\columnwidth}{!}{
  \begin{tabular}{cccc}
    \toprule
    Dataset & \#documents & \#queries & doc/query  \\
    \midrule
    ChemLit-QA & 1,528 & 1,025 & 6.00  \\
    SciFact & 5,183 & 300 & 1.13  \\
    LitSearch & 64,183 & 597 & 1.07  \\
    \bottomrule
  \end{tabular}}
  \vspace{-0.3cm}
\end{table}

We also employ three base retrievers through their official Hugging Face repositories:  
\begin{itemize}[leftmargin=10pt, itemsep=2pt]  
    \item \textbf{SPECTER2:} https://huggingface.co/allenai/specter2\_base  
    \item \textbf{Contriever-MS:} \href{https://huggingface.co/facebook/contriever-msmarco}{https://huggingface.co/facebook/ \\ contriever-msmarco}
    \item \textbf{Qwen3-Embedding-8B:} https://huggingface.co/Qwen/Qwen3-Embedding-8B  
\end{itemize}  
The parameter counts are approximately 110M for both SPECTER2 and Contriever-MS, and 8B for Qwen3-Embedding-8B.

\subsubsection{\textbf{Implementation of baselines}}
All baselines are implemented using publicly available code from the respective authors, with configurations and hyperparameters strictly following the original documentation.  
For BERT-QE \cite{zheng2020bertqe}, we set the number of top-ranked documents to 10.  
For LADR \cite{LADR}, we use the LADR-adaptive with no time constraints.  
For HyDE \cite{HyDE}, we follow the authors' prompts: \texttt{"Please write a scientific paper passage to answer the question"} for ChemLit-QA and LitSearch, and \texttt{"Please write a scientific paper passage to support/refute the claim"} for SciFact.  
For GRF \cite{grf}, ToTER \cite{toter}, and SemRank \cite{semrank}, we generate up to 50 semantic concepts, matching the number of candidates used in \method.  
For the pre-defined entity sets of ToTER and SemRank, we adopt the Microsoft Academic Graph \cite{wang2020microsoft}, as done in their original work.  

\subsubsection{\textbf{Implementation of PairSem}}
For agglomerative clustering in \S\ref{Sec:4.1.2}, we use the Scikit-learn library\footnote{https://scikit-learn.org/stable/modules/generated/sklearn.cluster.AgglomerativeClustering} \cite{pedregosa2011scikit}, setting the maximum cluster size to 20.  
For training the entity/aspect predictors in \S\ref{Sec:4.1.4}, the learning rate is selected from [1e-5, 1e-4] and the weight decay is selected from [0, 1e-6].  
For all LLM calls in baselines and \method, we use the OpenAI BatchAPI\footnote{https://platform.openai.com/docs/guides/batch} to reduce computation time and cost.  

\begin{table}[t!]
\caption{Evaluation on entity/aspect predictors.}
\vspace{-0.3cm}
\label{tab:entity_aspect_predictor}
\resizebox{1\columnwidth}{!}{
\begin{tabular}{lcccc}
\toprule
\multirow{2}{*}{Base Retriever} & \multicolumn{2}{c}{\textbf{Entity predictor}} & \multicolumn{2}{c}{\textbf{Aspect predictor}} \\
\cmidrule(lr){2-3} \cmidrule(lr){4-5}
 & P@10 & \#Entities & P@10 & \#Aspects \\
\midrule
\multicolumn{5}{l}{\textbf{\textit{ChemLit-QA}}} \\
SPECTER2 & 0.799 & \rdelim{\}}{3}{1cm}[3,352] & 0.790 & \rdelim{\}}{3}{1cm}[2,813] \\
Contriever-MS & 0.830 &  & 0.791 &  \\
Qwen3-Embedding-8B & 0.913 &  & 0.945 &  \\
\midrule
\multicolumn{5}{l}{\textbf{\textit{SciFact}}} \\
SPECTER2 & 0.797 &  \rdelim{\}}{3}{1cm}[17,697] & 0.801 &  \rdelim{\}}{3}{1cm}[5,312] \\
Contriever-MS & 0.800 &  & 0.818 &  \\
Qwen3-Embedding-8B & 0.843 &  & 0.893 &  \\
\midrule
\multicolumn{5}{l}{\textbf{\textit{LitSearch}}} \\
SPECTER2 & 0.713 & \rdelim{\}}{3}{1cm}[37,808] & 0.733 & \rdelim{\}}{3}{1cm}[6,540] \\
Contriever-MS & 0.730 &  & 0.771 & \\
Qwen3-Embedding-8B & 0.831 &  & 0.875 &  \\
\bottomrule
\end{tabular}}
\end{table}

\subsection{Supplementary Results}\label{sec:appenresult}
\subsubsection{\textbf{Performance of entity/aspect predictors}}
Table~\ref{tab:entity_aspect_predictor} presents the classification performance of the entity and aspect predictors (\S\ref{Sec:4.1.4}).
Each predictor is implemented as an $L$-layer MLP built on top of the embeddings produced by the base retriever ($L=2$ for SPECTER2 and Contriever-MS, and $L=5$ for Qwen3-Embedding-8B).
We report Precision@10 (P@10) over the corpus, which measures the proportion of Top-10 predicted entities or aspects that are actually in the semantic pairs generated by LLMs for each document.
`\#Entities' and `\#Aspects' denote the number of classes in our entity/aspect sets, after synonym merging.
Both predictors achieve high precision across datasets, demonstrating that the entity and aspect information is effectively captured within our framework.
Note that entity/aspect predictors are utilized to generate semantic pairs without LLMs for \methodfast (\S\ref{sec:4.2.1}), and to compute the entity-entity similarity $sim_{\text{entity}}(q,d)$ (\S\ref{Sec:4.2.2}).

\subsubsection{\textbf{Hyperparameter study}} 
\begin{figure}[t]
  \centering
  \includegraphics[width=1\linewidth]{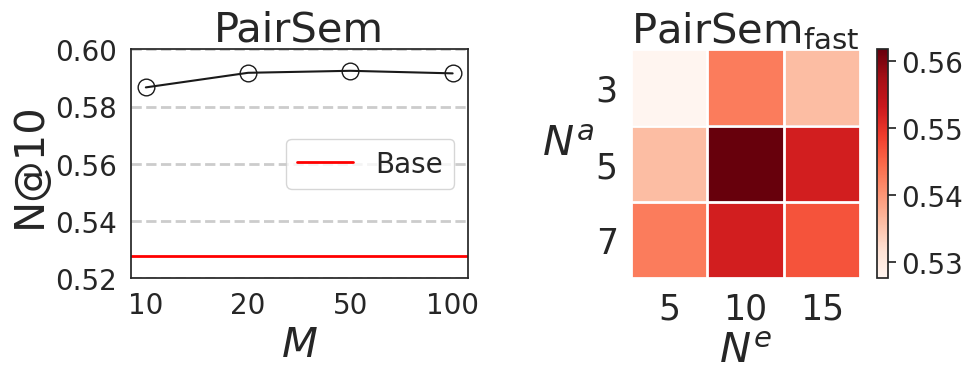}
  \vspace{-0.5cm}
  \caption{Hyperparameter study on \method and \methodfast.}
  \vspace{-0.2cm}
\label{fig:hyper}
\end{figure}
Figure~\ref{fig:hyper} presents the hyperparameter study of \method and \methodfast on ChemLit-QA, using SPECTER2 as the base retriever.
For \method, we vary $M$, the number of candidate entities and aspects used for candidate-augmented pair generation (\S\ref{Sec:4.1.3}).
For \methodfast, we adjust $N^e$ and $N^a$, the numbers of top entities and aspects estimated by the entity/aspect predictor, to construct semantic pairs of queries without invoking an LLM (\S\ref{sec:4.2.1}).
The results demonstrate that both methods consistently improve retrieval performance over the base retriever (N@10=0.5279) across all tested hyperparameter settings.

\begin{table*}[t!]
\centering
\caption{Detailed case study on ChemLit-QA dataset with SPECTER2 (extended from Figure~\ref{fig:intro}). 
For SemRank \cite{semrank}, we highlighted overlapped semantic entities in \sethlcolor{mediumgrey}\hl{Grey}. For \method, we highlighted relevant entities in \sethlcolor{pastelred}\hl{Red} and associated aspects in \sethlcolor{pastelblue}\hl{Blue}}
\label{tab:casestudy}
\setlength{\tabcolsep}{3pt}
\resizebox{\textwidth}{!}{
\begin{tabular}{p{0.08\textwidth}|p{0.45\textwidth}|p{0.08\textwidth}|p{0.45\textwidth}}
\toprule
Query & \multicolumn{3}{l}{How does the adsorption of perfluorinated acids onto the metal-organic framework's active sites lead to the detection of PFAS in water?} \\
\cline{2-4}
& \multicolumn{3}{l}{\parbox[t]{1\textwidth}{
\textbf{Generated semantic pairs (by PairSem):} \\
(\sethlcolor{pastelred}\hl{perfluorinated acid}, \sethlcolor{pastelblue}\hl{adsorption}), (\sethlcolor{pastelred}\hl{metal-organic framework}, \sethlcolor{pastelblue}\hl{active site}),
(\sethlcolor{pastelred}\hl{PFAS}, \sethlcolor{pastelblue}\hl{detection}),
(\sethlcolor{pastelred}\hl{water}, \sethlcolor{pastelblue}\hl{medium})
}} \\
\midrule
Doc A \newline (relevant) \newline \newline rank:~19 & 
... by fully immersing the \textbf{metal-organic framework (MOF)}-covered channel into contaminated \textbf{water}. The device resistance showed to decrease significantly upon immersion, even when as little as 10 fM of the \textbf{perfluorinated acid} was present. With careful dilution of the analyte solution, we found a detection limit of our \textbf{MOF-based sensor} to be as low as 5 fM. Such sensitivity could be attributed to the high surface area of our \textbf{MOF} films, as well as the chemical affinity of the metal \textbf{active sites} towards the \textbf{perfluorinated acid}. Furthermore, we tested 10 different devices and showed a linear correlation (R$^2$ > 0.968) between the conductance and the logarithm of \textbf{PFAS} concentration indicative of the reliability of the \textbf{MOF} film approach ... 
As previously mentioned, the \textbf{PFAS} sensing is based on the adsorption of the \textbf{PFAS} molecules onto the \textbf{metal-organic framework}'s sites which further allowed our sensors ... &

Doc B \newline (irrelevant) \newline \newline rank:~10 & 
... observed oxidation of the \textbf{MOF film} also showed to be non-destructive for the \textbf{MOF structure}. … the structure of the \textbf{MOF} remained intact after dipping in \textbf{PFAS} and washing. The elemental composition of the \textbf{MOF film} was found to remain unchanged as evidenced from the post-sensing XPS spectra. The structural stability of \textbf{metal-organic framework} towards the efficient release of \textbf{PFAS molecules} was rationalized by the fact that the key interactions between the \textbf{MOF} and the analyte are due to surface electrostatics that are electronically detectable yet reversible especially at low \textbf{PFAS} concentrations. … We then tested the sensitivity of our \textbf{MOF films} towards \textbf{PFAS} using electrochemical impedance spectroscopy (EIS) as a means to monitor the change in channel conductance. The simple device architecture of our studied \textbf{MOF-based sensor} is shown in Figure a. We employ a 2-terminal device layout ...  \\
\midrule

\textbf{SemRank} \cite{semrank} \newline \newline rank:~15 & 
\textbf{Semantic entities:} \sethlcolor{mediumgrey}\hl{metal-organic framework}, \sethlcolor{mediumgrey}\hl{perfluorinated acid}, \sethlcolor{mediumgrey}\hl{biochemical analyte}, \sethlcolor{mediumgrey}\hl{polymer adsorption}, \sethlcolor{mediumgrey}\hl{active site}, \sethlcolor{mediumgrey}\hl{electrostatic discharge materials}, contamination water, groundwater pollution, immersion, electron affinity &

\textbf{SemRank} \cite{semrank} \newline \newline rank:~12 & 
\textbf{Semantic entities:} \sethlcolor{mediumgrey}\hl{metal-organic framework}, \sethlcolor{mediumgrey}\hl{perfluorinated acid}, \sethlcolor{mediumgrey}\hl{biochemical analyte}, \sethlcolor{mediumgrey}\hl{polymer adsorption}, \sethlcolor{mediumgrey}\hl{active site}, \sethlcolor{mediumgrey}\hl{electrostatic discharge materials}, oxidation activity, structural stability \\
\midrule

\textbf{PairSem} (ours) \newline \newline rank:~4 & 
\textbf{Semantic pairs:}
(\sethlcolor{pastelred}\hl{metal-organic framework}, \sethlcolor{pastelblue}\hl{active site}), \newline
(\sethlcolor{pastelred}\hl{metal-organic framework}, reusability), 
(\sethlcolor{pastelred}\hl{PFAS}, \sethlcolor{pastelblue}\hl{detection}), \newline
(\sethlcolor{pastelred}\hl{PFAS}, concentration), 
(\sethlcolor{pastelred}\hl{perfluorinated acid}, \sethlcolor{pastelblue}\hl{adsorption}), \newline
(\sethlcolor{pastelred}\hl{water}, \sethlcolor{pastelblue}\hl{medium}), ...
&

\textbf{PairSem} (ours) \newline \newline rank:~26 & 
\textbf{Semantic pairs:}
(\sethlcolor{pastelred}\hl{metal-organic framework}, oxidation), \newline
(\sethlcolor{pastelred}\hl{metal-organic framework}, structural stability), \newline
(\sethlcolor{pastelred}\hl{metal-organic framework}, sensitivity),
(\sethlcolor{pastelred}\hl{PFAS}, release),\newline
(analyte, binding free energy), ...
\\

\bottomrule
\end{tabular}
}
\end{table*}

\subsubsection{\textbf{Case study}}
Table~\ref{tab:casestudy} presents a detailed case study on the ChemLit-QA dataset using SPECTER2 as the base retriever.
We observe that \method effectively identifies meaningful semantic pairs in the query, where entities are explicitly linked with their associated aspects.
In contrast, a flat list of semantic entities (as in SemRank) struggles to distinguish the two documents, since most identified entities are shared across both.
By coupling entities with aspects, \method can successfully differentiate the relevance of documents.
For instance, although both documents mention \texttt{"PFAS"}, Doc A focuses on the \texttt{"detection"} of \texttt{"PFAS"}, whereas Doc B discusses the \texttt{"release"} of \texttt{"PFAS"}.
This demonstrates that \method enables finer-grained semantic understanding, capturing aspect-level differences that are overlooked by entity-based methods, thereby yielding more accurate relevance estimation.

\end{document}